\documentclass[twocolumn,pra,showpacs,superscriptaddress,amssymb,amsmath,amsmath]{revtex4-1}
\usepackage{graphicx}
\usepackage{epstopdf}
\usepackage{bm}
\usepackage{hyperref}
\usepackage{comment}

\hypersetup{%
   pdfpagemode=None, 
   pdfstartpage=1,
   pdfmenubar=true,
   pdftoolbar=true,
   colorlinks = true,
   linkcolor=blue,
   citecolor=blue,
   urlcolor=blue,
   bookmarksopen=false
 }

\newcommand{\be}{\begin{equation}}
\newcommand{\ee}{\end{equation}}

\begin{document}
\title{\textit{Ab initio} electronic structure of the Sr$_2^+$ molecular ion}

\author{Micha\l~\'Smia{\l}kowski}
\affiliation{Faculty of Physics, University of Warsaw, Pasteura 5, 02-093 Warsaw, Poland}
\affiliation{Faculty of Chemistry, University of Warsaw, Pasteura 1, 02-093 Warsaw, Poland}
\author{Tatiana Korona}
\affiliation{Faculty of Chemistry, University of Warsaw, Pasteura 1, 02-093 Warsaw, Poland}
\author{Micha\l~Tomza}
\email{michal.tomza@fuw.edu.pl}
\affiliation{Faculty of Physics, University of Warsaw, Pasteura 5, 02-093 Warsaw, Poland}

\date{\today}

\begin{abstract}

Molecular ions formed in cold hybrid ion-atom experiments may find interesting applications ranging from precision measurements to controlled chemical reactions. Here, we investigate electronic structure of the Sr$_2^+$ molecular ion, which may be produced by photoassociation of laser-cooled Sr$^+$ ions immersed into an ultracold gas of Sr atoms or by ionization of ultracold Sr$_2$ molecules. Using \textit{ab initio} electronic structure methods, such as the coupled cluster and configuration interaction ones with small-core relativistic energy-consistent pseudopotentials and large Gaussian basis sets, we calculate potential energy curves for the ground and 41 excited electronic states, and electric dipole transition moments between them. We show that alkaline-earth molecular ions despite of their apparently simple structure with three valence electrons only are challenging for state-of-the-art quantum chemistry methods due to their multireference nature and high density of states. Finally, we calculate and analyze Franck-Condon factors governing the photoionization of ground-state Sr$_2$ molecules into $^2\Sigma^+_u$ and $^2\Sigma^+_g$ states of Sr$_2^+$ molecular ions. The present results may be useful for studying and guiding formation and spectroscopy of cold Sr$_2^+$ molecular ions. 

\end{abstract}

\maketitle

\section{Introduction}

Since the onset of the twenty-first century the ultracold matter experiments have increasingly directed towards the formation of molecules from laser-cooled atoms or direct cooling molecules from higher temperatures~\cite{CarrNJP09,QuemenerCR12,BohnS17}. These investigations have been motivated by both fundamental and practical dimensions with possible applications ranging from controlled chemical reactions~\cite{KremsPCCP08,TomzaPRL15,BalakrishnanJCP16} and precision measurements~\cite{DeMilleScience17,SafronovaRMP18} to quantum simulation~\cite{MicheliNatPhys06,BaranovCR12} and quantum computation~\cite{DeMillePRL02,Albert2019}. Following the successful experiments with ultracold molecules, increasingly more trapped molecular ions have been studied at low temperatures~\cite{WillitschIRPC12,TomzaRMP19}. Molecular ions can be either produced at higher temperatures and subsequently cooled down~\cite{TongPRL10,StaanumNP10} or formed from ultracold ion-atom mixtures via association~\cite{SullivanPCCP11,TomzaPRA15a} or from ulracold molecules via ionization~\cite{JyothiPRL16,SchmidPRL18}. 

High controllability of ultracold molecules is enhanced even further in case of molecular ions, as they exhibit significantly longer-range interactions than neutral species, and can be manipulated and detected on a single particle level in ion traps~\cite{WinelandRMP13,TomzaRMP19}. This can allow for high precision spectroscopy~\cite{LohScience13,GermannNatPhys14,KajitaPRA14} and measurements of chemical reaction rates with particle densities of molecular ions much smaller and better controlled than with neutral molecules~\cite{PuriNC19,Dorfler2019}. Moreover, the charged products of chemical reactions can be trapped thus opening the way for measuring product-state distributions and state-to-state reaction rates~\cite{ChangScience13,PuriScience17,KilajNC18}. Sympathetically cooled molecular ions such N$_2^+$~\cite{HallPRL12} and OH$^-$~\cite{DeiglmayrPRA12} were already immersed into ultracold Rb atoms, while BaCl$^+$ molecular ions were sympathetically cooled down by collisions with ultracold Ca atoms~\cite{RellergertNature13,StoecklinNP16}.

Enormous successes of experiments with ultracold alkali and alkaline-earth atoms and ions result from their electronic structures favorable for laser-cooling. Therefore, first cold ion-atom mixtures unsurprisingly employed alkaline-earth ions and alkali or alkaline-earth atoms~\cite{MakarovPRA03,GrierPRL09,HallMP13a,ZipkesNature10,HazePRA13,HallPRL11,RaviNatCommun12,HartePRL12,SmithAPB14,MeirPRL16,JogerPRA17}. The radiative formation of RbCa$^+$~\cite{HallPRL11, HallMP13b}, RbBa$^+$~\cite{HallMP13a}, CaYb$^+$~\cite{RellergertPRL11} and CaBa$^+$~\cite{SullivanPCCP11} molecular ions was already observed in cold collisions between respective ions and atoms. Rb$_2^+$~\cite{JyothiPRL16} and Ca$_2^+$~\cite{SullivanPCCP11} molecular ions were formed by the photoionization of ultracold molecules. Both approaches may be used for the formation of the Sr$_2^+$ molecular ion studied in this paper. 

Neutral alkaline-earth dimers have been spectroscopically studied in a number of experiments, including  Be$_2$~\cite{MerrittPCCP08,MerrittS09}, Mg$_2$~\cite{BalfourCJP70,McCaffreyJCP88,KnockelJCP13}, Ca$_2$~\cite{BalfourCJP75,vidalJCP80,bondybeyCPL84,AllardPRA02,AllardEPJD05}, Sr$_2$~\cite{BergemanJCP80,GerberJCP84,SteinPRA08,SteinEPJD10,SteinEPJD11}, Ba$_2$~\cite{boutouZP97,lebeaultJMS98}.
They also received considerable attention as the subject of theoretical studies~\cite{JonesJCP79,PartridgeJCP90,AlloucheCP95,CzuchajTCA03,BusseryPRA03,BusseryPRA05,LeeJPCA05,BusseryJCP06,BusseryMP06,PatkowskiJPCA07,PatkowskiS09,
BouissouJCP10,FromagerPRA10,heavenCPL11,LiJPCA11,SchaferPRA07,
yangTCA12,KerkinesJCP12,LiJPCA11,AmitayFD11,MeshkovJCP14,SharmaJCP14,WeiCPL15,KalemosJCP16,nasiriCTC17,StephenMP19,LesiukJCTC19}, including several works on Sr$_2$~\cite{BoutassettaPRA96,WangJPCA00,BeloyPRA11,czuchajCPL03,KotochigovaJCP08,KochPRA08b,SkomorowskiPRA12,SkomorowskiJCP12}.
Recently, ground breaking experiments with an ultracold gas of Sr$_2$ molecules in an optical lattice have been realized~\cite{ZelevinskyPRL06,ReinaudiPRL12,StellmerPRL12,CiameiPRA17} to study both high precision spectroscopy~\cite{ZelevinskyPRL08,McGuyerPRL13,McGuyerNP15,McGuyerPRL15,KondovNP19} and controlled photodissociation~\cite{mcdonaldN16,McDonaldPRL18,KondovPRL18}. Despite having only four valence electrons, the Sr$_2$ molecule turned out to be a challenging system for an accurate theoretical description of its electronic structure. An initial disagreement between relativistic calculations~\cite{KotochigovaJCP08} and experimental results~\cite{SteinPRA08} for excited electronic states was later resolved using higher level calculations~\cite{SkomorowskiJCP12}.

Surprisingly, there are only a few experimental or theoretical studies of alkaline-earth molecular ions. The dissociation energies were experimentally measured for Be$_2^+$~\cite{MerrittS09,AntonovJCP10} and Sr$_2^+$~\cite{dugourdCPL92} using photoionization of neutral dimers. The ground electronic state was theoretically investigated for all alkaline-earth molecular ions~\cite{ABC+,LiMP13}. The lowest excited states were studied for Be$_2^+$~\cite{BanerjeeCPL10}, Mg$_2^+$~\cite{StevensJCP77,AlharzaliJPB18}, and Ca$_2^+$~\cite{LiuPRA78,BanerjeeCPL12}. To the best of our knowledge, excited states of the Sr$_2^+$ molecular ion have not yet been investigated theoretically or experimentally. 

Here, we fill this gap and calculate the electronic structure of the Sr$_2^+$ molecular ion using \textit{ab initio} methods of quantum chemistry. We characterize and benchmark two lowest electronic states of the Sr$_2^+$ molecular ion with a range of computational techniques including hierarchy of configuration interaction and coupled cluster methods. Next, we use multireference configuration interaction method restricted to single and double exctiations to obtain potential energy curves for 41 excited electronic states and electric dipole transition moments between them. 11 excited states are also obtained with the coupled cluster method restricted to single, double, and noniterative triple excitations. We show that the Sr$_2^+$ molecular ion despite of its apparently simple structure with three valence electrons only is challenging for state-of-the-art quantum chemistry methods due to its multireference nature and high density of states. Finally, we provide Franck-Condon factors governing the photoionization of ground-state Sr$_2$ molecules into $^2\Sigma^+_u$ and $^2\Sigma^+_g$ states of Sr$_2^+$ molecular ions.  The Sr$_2^+$ molecular ion is an interesting system for theoretical studies because it can be produced and investigated in modern experiments with ultracold Sr$_2$ molecules or with mixtures of Sr$^+$ ions immersed into ultracold Sr atoms. Thus, our results may be useful for guiding future formation and spectroscopic measurements of the Sr$_2^+$ molecular ion in the ground and excited electronic states. 

The paper has the following structure. Section~\ref{sec:theory} concerns theoretical methods used in the \textit{ab initio} electronic structure calculations. Section~\ref{sec:results} presents and discusses the results obtained for the ground and excited electronic states of the Sr$_2^+$ molecular ion. Section~\ref{sec:summary} summarizes the paper and points to applications and extensions of the presented results.

\section{Computational details}
\label{sec:theory}

The Sr$_2^+$ molecular ion is composed of a closed-shell strontium atom with two valence electrons interacting with an open-shell strontium ion with one valence electron. Therefore, the resulting system has three valence electrons with a doublet multiplicity of the ground electronic state, while excited states can be either doublets or quartets. Excited electronic states correlate with atomic thresholds resulting from exciting the Sr$^+$ion or Sr atom, or both of them (see Tables~\ref{tab:atomic} and \ref{tab:thresholds}).  

Here, to calculate potential energy curves in the Born-Oppenheimer approximation we adopt the computational scheme successfully applied to the ground and excited electronic states of the SrYb and Sr$_2$ molecules~\cite{TomzaPCCP11,SkomorowskiJCP12}, and LiYb$^+$ molecular ion~\cite{TomzaPRA15a}. All interaction and excitation energies, as well as electric dipole transition moments, are obtained with the multireference configuration interaction method restricted to single and double excitations (MRCISD) using orbitals optimized with the multi-configurational self-consistent field method (MCSCF)~\cite{MRCC} with a large active space composed of all molecular orbitals created from the $5s$, $5p$, $5d$, and $6s$ orbitals of both the Sr atom and the Sr$^+$ ion. The ground state and the lowest energetic excited electronic states in each irreducible representation of the $D_{2h}$ point group are additionally computed with the spin-restricted open-shell coupled cluster method restricted to single, double, and non-iterative triple excitations (RCCSD(T))~\cite{PurvisJCP82,KnowlesJCP93}. For comparison, we also present results for the $X^2\Sigma_u^+$ ground and the $^2\Sigma_g^+$ lowest-energetic excited state with the following methods: open-shell M$\o$ller-Plesset theory (RMP2)~\cite{MP2}, the spin-restricted open-shell coupled cluster method restricted to single and double excitations (RCCSD)~\cite{KnowlesJCP93}, MRCISD with the Davidson correction (MRCISD+Q)~\cite{MRCC}, and configuration interaction method restricted to single and double excitations (CISD)~\cite{MRCC}.
The interaction energy,~$E_\text{int}$, is computed as the difference between the energy of the dimer in a given state,~$E_{\text{Sr}_2^+(^{2S+1}\Lambda)}$, and the energies of the atom,~$E_{\text{Sr}(^{2s+1}L)}$, and ion,~$E_{\text{Sr}^+(^{2s'+1}L')}$, in the electronic states corresponding to the dissociation limit of the dimer in the~$^{2S+1}\Lambda$~state,
\begin{equation}
E_\text{int}=E_{\text{Sr}_2^+(^{2S+1}\Lambda)}-E_{Sr(^{2s+1}L)}-E_{Sr^+(^{2s'+1}L')}\,.
\end{equation}
All results are obtained by applying the Boys-Bernardi counterpoise correction method~\cite{BoysMP70}. All electronic structure calculations are performed with the \textsc{Molpro} package of \textit{ab initio} programs \cite{Molpro,MOLPRO-WIREs}.

\begin{table}[tb]
\caption{Atomic excitation energies of the Sr~atom and the Sr$^+$~ion, and the ionization potential of the Sr~atom calculated with the MRCISD and CCSD(T) methods compared with experimental data. All energies are in cm$^{-1}$.\label{tab:atomic}} 
\begin{ruledtabular}
\begin{tabular}{lrrr}
Electronic transition    & MRCI & CC & Exp.~\cite{NIST} \\ 
\colrule
Sr$(^1S$ $\to$ $^3P)$   &  14139      &    14641    & 14703 \\
Sr$(^1S$ $\to$ $^3D)$   &  19071  &  18745 &  18254 \\
Sr$^+(^2S$ $\to$ $^2D)$   &  15553  &   15371 & 14743  \\
Sr$^+(^2S$ $\to$ $^2D)$   & 23839  & 24152 & 24249   \\
Sr$(^1S)$ $\to$ Sr$^+(^2S)$ & 44824 & 45820 & 45932 \\
\end{tabular}
\end{ruledtabular}
\end{table}

Scalar relativistic effects are included by replacing 28~inner-shell electrons in both the Sr atom and the Sr$^+$ ion with the small-core, fully relativistic, energy-consistent pseudopotential ECP28MDF~\cite{LimJCP05} from the Stuttgart  library. Thus, in the present study the Sr$_2^+$ molecular ion is treated as a system of effectively correlated 19~electrons. The additional advantage of employing pseudopotentials is the possibility to use larger basis sets to describe the valence electrons, while the inner-shell electron density is reproduced with the accuracy of high-quality atomic calculations. To this end, we utilize the large [14s11p6d5f4g] basis set proposed in Refs.~\cite{TomzaPCCP11,SkomorowskiJCP12}, augmented by the set of the [3s3p2d2f1g] bond functions~\cite{midbond}. The electric dipole transition moments are computed using respective MRCISD wave functions.

\begin{table*}[tb!]
\caption{Asymptotic energies (in cm$^{-1}$) and molecular states arising from different states of the strontium atom and ion. MRCI excitation energies  are averaged out over separate calculations for different molecular symmetries due to the size inconsistency of the MRCI method. \label{tab:thresholds}} 
\begin{ruledtabular}
\begin{tabular}{lrrrl}
Asymptote    & MRCI energy & CC energy & Exp.~energy~\cite{NIST} &  Molecular states \\ 
\colrule
Sr$^+(^2S)$~+~Sr$(^1S)$   &  0      &    0    & 0  & $^2\Sigma_g$, $^2\Sigma_u^+$  \\
Sr$^+(^2S)$~+~Sr$(^3P)$   &  14100  &  14612 &  14703 &  $^2\Sigma_g$,$^2\Sigma_u^+$, $^2\Pi_g$,  $^2\Pi_u$, $^4\Sigma_g^+$, $^4\Sigma_u^+$, \\
& & & & $^4\Pi_g$, $^4\Pi_u$ \\
Sr$^+(^2D)$~+~Sr$(^1S)$   &  16850  &   15843 & 14743 & $^2\Sigma_g^+$, $^2\Sigma_u^+$, $^2\Pi_g$, $^2\Pi_u$, $^2\Delta_g$, $^2\Delta_u$,    \\
Sr$^+(^2S)$~+~Sr$(^3D)$   & 19100  & 18770 & 18254 &  $^2\Sigma_g$, $^2\Sigma_u^+$, $^2\Pi_g$, $^2\Pi_u$, $^2\Delta_g$, $^2\Delta_u$,    \\
                    &        &       &         & $^4\Sigma_g^+$, $^4\Sigma_u$, $^4\Pi_g$, $^4\Pi_u$, $^4\Delta_g$, $^4\Delta_u$ \\
Sr$^+(^2S)$~+~Sr$(^1D)$   & 20950  & - & 20150  &$^2\Sigma_g^+$, $^2\Sigma_u$, $^2\Pi_g$, $^2\Pi_u$, $^2\Delta_g$, $^2\Delta_u$,    \\
Sr$^+(^2S)$~+~Sr$(^1P)$   & 23000  & - &  21698 & $^2\Sigma_g^+$, $^2\Sigma_u$, $^2\Pi_g$, $^2\Pi_u$    \\
Sr$^+(^2P)$~+~Sr$(^1S)$   & 24100 & - & 24249 &$^2\Sigma_g^+$, $^2\Sigma_u$, $^2\Pi_g$, $^2\Pi_u$    \\
\end{tabular}
\end{ruledtabular}
\end{table*}

To evaluate the ability of the employed \textit{ab initio} approaches to reproduce experimental results, we first compared the theoretical excitation energies of a strontium atom and a strontium ion obtained in this study to the experimental values~\cite{NIST} (see Table~\ref{tab:atomic}). The CCSD(T) excitation energies of Sr and Sr$^+$ to the lowest $P$~state agree with the experimental results within 0.5\%, while the excitations to the lowest $D$~state agree within 3\% for the neutral atom and within 5\% for the ion. Additionally, the CCSD(T) ionization energy of the ground-state Sr atom to the ground-state ion is within 0.25\% from the experimental value. The MRCISD method yields slightly less accurate results: the errors of the excitation energies for the neutral strontium atom amount to 3.8\% for the $P$~state and to 4.5\% for the $D$~state. As for the strontium ion, the errors are equal to 1.7\% for the $P$ state and to 5.6\% for the $D$~state excitation. Additionally, the MRCISD ionization energy is within 2.4\% from the experimental value. Therefore, for both species, the excitations into the $D$ states are less accurately described by the basis set we employ than the $P$~ones, with the Sr$^+$ ion $D$~state exhibiting the lowest accuracy. However, further basis extension with additional $d$~orbitals does not improve the accuracy meaningfully. The static electric dipole polarizabilities of the Sr atom and Sr$^+$ ion obtained with the CCSD(T) method are 199.2$\,$a.u.~and 92.0$\,$a.u. and agree within 1\% with experimental values of 197.1(2)$\,$a.u.~and 91.3(9)$\,$a.u.~\cite{MitroyJPB10}, respectively.

\section{Results}
\label{sec:results}

\subsection{Potential energy curves}

We have calculated potential energy curves~(PECs) and their spectroscopic characteristics for the doublet and quartet gerade and ungerade electronic states of the Sr$_2^+$ molecular ion corresponding to the first seven lowest dissociation limits, Sr$^+$($^2$S)+Sr($^1S$), Sr$^+$($^1S$)+Sr($^3P$), Sr$^+$($^2D$)+Sr($^1S$), Sr$^+$($^2S$)+Sr($^3D$), Sr$^+$($^2S$)+Sr($^1D$), Sr$^+$($^2S$)+Sr($^1P$), and Sr$^+$($^2P$)+Sr($^1S$), that gives in total 42~electronic states. Molecular electronic states arising from atomic asymptotes are collected in Table~\ref{tab:thresholds}.

\begin{figure}[tb]
\begin{center}
\includegraphics[width=0.98\columnwidth]{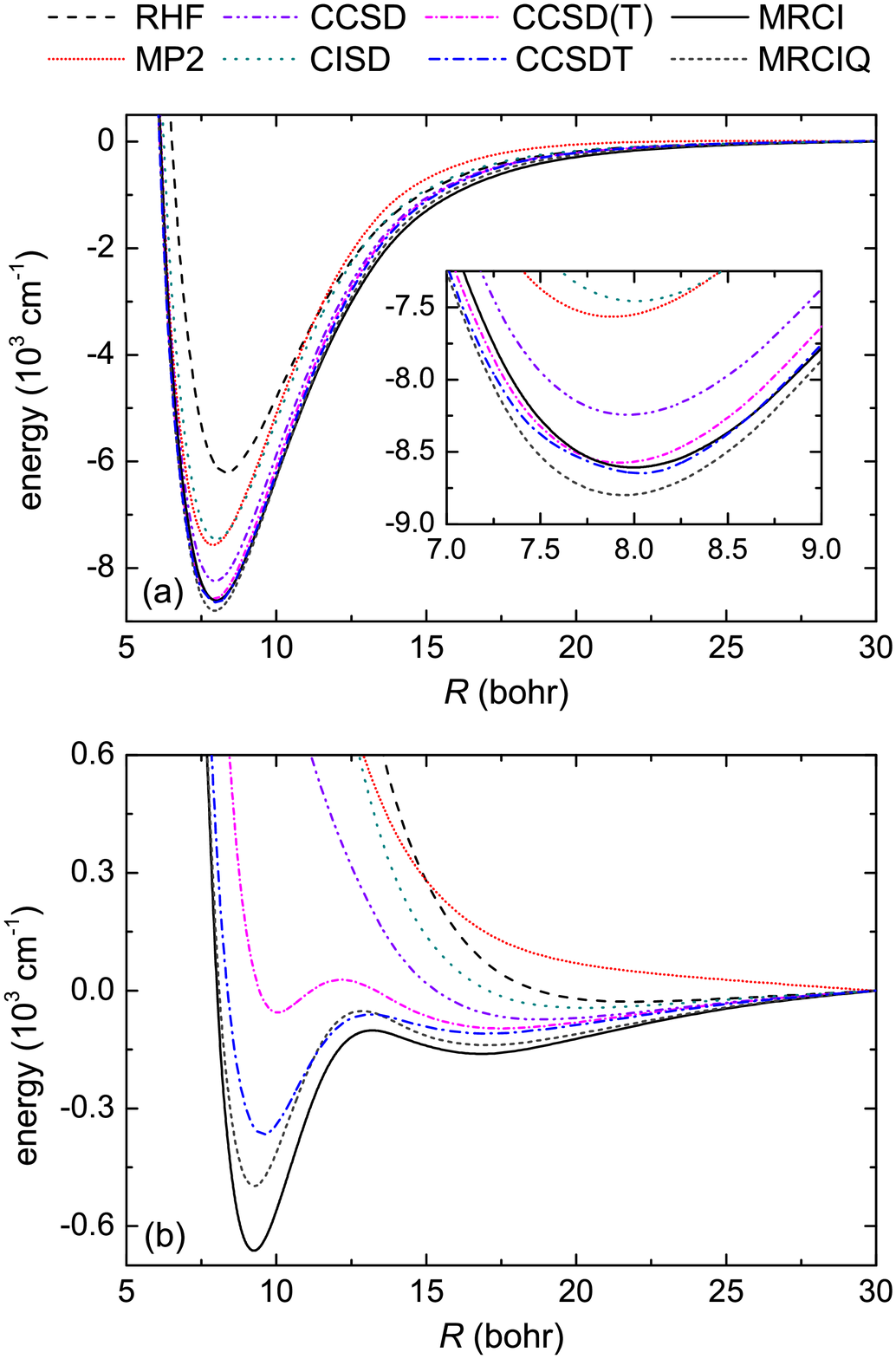}
\end{center}
\caption{Non-relativistic potential energy curves for (a)~the $X^2\Sigma^+_u$ ground and (b)~the $1^2\Sigma^+_g$ lowest-energetic excited electronic states of the Sr$_2^+$ molecular ion calculated at different levels of theory.}
\label{fig:X}
\end{figure}

The $X^2\Sigma_u^+$ ground state potential energy curve is calculated by a range of \textit{ab initio} methods and presented in Fig.~\ref{fig:X}, while corresponding spectroscopic parameters are collected in Table~\ref{tab:su}. In the naive molecular orbital theory picture, the Sr$_2^+$ molecular ion in the $X^2\Sigma^+_u$ ground state can be considered as a bound molecule because its valence molecular configuration, $\sigma_g^2\sigma_u^{*1}$, has one more bonding than antibonding electron resulting in a bond of order one-half. Therefore, all employed electronic structure methods describe the ground-state Sr$_2^+$ molecular ion qualitatively correctly, albeit to a varying degree of accuracy.  Compared to the experimental value of dissociation energy at $8800(130)$~cm$^{-1}$~\cite{dugourdCPL92}, the MRCISD method with the Davidson correction returns the most accurate result, followed closely by CCSDT~\cite{MRCC}, which includes a full triple excitations in the CC scheme, and the regular MRCISD and CCSD(T) methods. Other approaches are less accurate: the CCSD method exhibits an error larger than 5\%, both MP2 and CISD methods 
give as much as 15\% error, while for restricted Hartree-Fock method the error increases further to 30\% . 

In contrast to the ground state of the Sr$_2^+$ molecular ion, the 1$^2\Sigma_g^+$ lowest-energetic excited electronic state in the naive molecular orbital theory picture has one more antibonding than bonding electrons in its valence molecular configuration, $\sigma_g^1\sigma_u^{*2}$. For this reason, it is a challenging system for several \textit{ab initio} methods, which we compare in Fig.~\ref{fig:X} and Table~\ref{tab:su}. For instance, it is interesting to note that the M$\o$ller-Plesset interaction energy curve lies above the restricted Hartree-Fock one for intermediate distances from around 15 to 30 bohrs. This behavior can be justified by the fact that both monomers (Sr and Sr$^+$) are single-reference species, and therefore are well described already on the MP2 level, while the dimer has a multireference character which can be seen in the large differences between single and multireference correlated methods. For the same reason, coupled cluster methods restricted to single and double excitations fail to describe the system with size consistency, while it is preserved in the RCCSD(T) and RCCSDT methods. Note parenthetically that the curves in Fig.~\ref{fig:X} are shifted to the same asymptotic limit for a better visibility.

\begin{table}[tb]
\caption{Spectroscopic characteristics of the Sr$_2^+$ molecular ion in the $X^2\Sigma^+_u$ ground and $1^2\Sigma^+_g$ lowest-excited electronic states: equilibrium bond length~$R_e$, well depth~$D_e$, harmonic constant~$\omega_e$, and rotational constant~$B_e$, calculated at different levels of theory. \label{tab:su}} 
\begin{ruledtabular}
\begin{tabular}{lrrrr}
Ref. & $R_e\,$(bohr) & $D_e\,$(cm$^{-1}$) &  $\omega_e\,$(cm$^{-1}$) & $B_e\,$(cm$^{-1}$) \\
\hline
\multicolumn{5}{c}{$X^2\Sigma^+_u$} \\
RHF & 8.33 & 6206 & 71.2 & 0.0198 \\
MP2 & 7.88 & 7569 & 80.1 & 0.0221  \\
CISD  & 8.00 & 7461 & 79.2 & 0.0215 \\
CCSD & 7.96 & 8245 & 79.5 & 0.0217  \\
CCSD(T)~\cite{ABC+} & 7.93 & 8576 &  79.9 & 0.0219 \\
CCSDT &  7.93 & 8703 & 80.1 & 0.0219 \\
MRCISD  & 7.99 & 8611 & 78.9 & 0.0215  \\
MRCISD+Q & 7.94 & 8801 & 80.2 & 0.0218 \\
Theo.~\cite{LiMP13} & 7.90 & 8745 & 80.7 & 0.022 \\
Exp.~\cite{dugourdCPL92} & - & 8800(130) & 86(3) & - \\
\hline
\multicolumn{5}{c}{$1^2\Sigma^+_g$} \\
RHF &  20.46 & 30 & 3.3 & 0.00328\\
MP2 &  \multicolumn{4}{c}{repulsive} \\
CISD  &  19.29 & 56 & 4.8 & 0.00369 \\
CCSD & 18.33 & 108 & 5.1 & 0.00409  \\
CCSD(T) & 10.03 & 78 & 23.7 & 0.01365 \\
2$^\text{nd}$ min. & 17.39 & 126 & 5.0 & 0.00455 \\
CCSDT & 9.55 & 359 & 29.5 & 0.01506 \\
2$^\text{nd}$ min. &  17.06 & 106 & 5.0 & 0.00472 \\
MRCISD  &  9.23 & 678 & 37.5 & 0.01612 \\
2$^\text{nd}$ min.  &  17.04 & 184 & 6.3 & 0.00473 \\
MRCISD+Q &  9.24 & 506 & 34.7 & 0.01608 \\
2$^\text{nd}$ min.  & 16.90 & 153 & 5.9 & 0.00481  \\
\end{tabular}
\end{ruledtabular}
\end{table}

The next significant feature of the 1$^2\Sigma_g^+$ curve is the presence of two minima: the short-range and the long-range ones. Similar double well structure for this electronic state was also theoretically predicted for the Be$_2^+$~\cite{BanerjeeCPL10} and Ca$_2^+$~\cite{BanerjeeCPL12} dimers. Less accurate and single-reference methods, like RHF, CISD and CCSD, detect only the latter minimum. The interactions at distances around the long-range minimum should be dominated by the induction component, however a more detailed analysis of this issue with e.g.~symmetry adapted perturbation theory~\cite{SAPT} is challenging due to the Sr$_2^+$ homonuclear structure and charge delocalization. The addition of non-iterative or full triples excitations in, respectively, the CCSD(T) and CCSDT methods results in the second minimum formation, which is further deepened by the MRCISD and MRCISD+Q methods. Thus, the existence of the short-range minimum arises from the multireference nature of the 1$^2\Sigma_g^+$ state at short distances in agreement with previous predictions for other alkaline-earth diatomic ions~\cite{BanerjeeCPL10,BanerjeeCPL12}. The analysis of the highest excitation amplitudes in the CCSD versus MRCISD methods shows that the two most important configuration functions (apart from the reference one) are strongly underestimated within the former approach. For instance, the double excitation from the highest occupied $\sigma^*_u$ orbital into the singly occupied $\sigma_g$ and the lowest unoccupied $\sigma_g$ is present in CCSD with the coefficient of about 0.3 for distance 8.5 bohr, while for MRCISD its contribution amounts to 0.8 (after applying intermediate normalization for MRCI). The situation is analogous for the second most important configuration which describes the excitation of both electrons from the same occupied $\sigma^*_u$ orbital into the lowest unoccupied $\sigma_g$ orbital (the CCSD coefficient of 0.2 is about two times too small when compared to the MRCISD one). These serious deficiencies of the CCSD model can explain the lack
\begin{figure}[hptb!]
\begin{center}
\includegraphics[width=0.96\columnwidth]{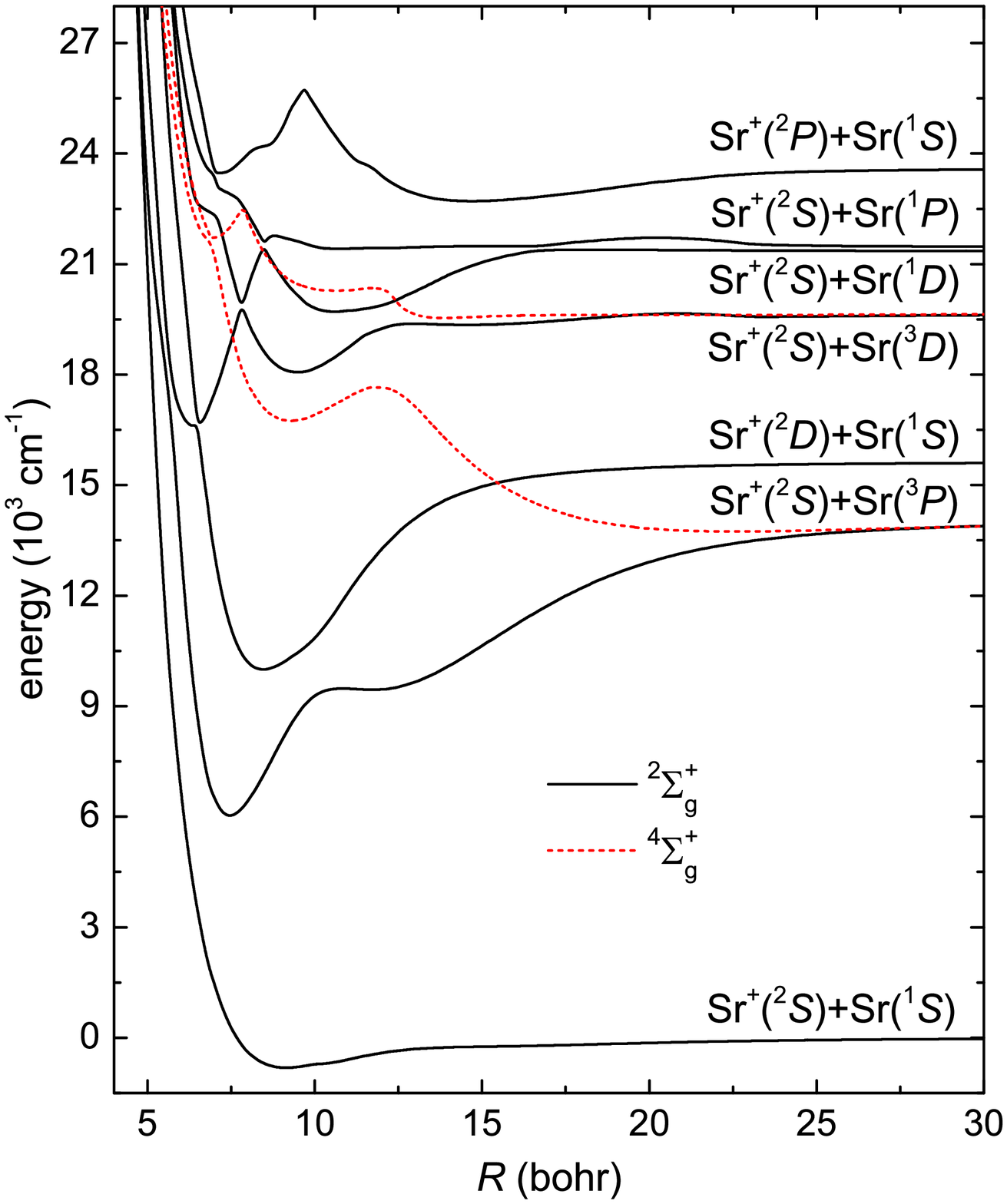}
\end{center}
\caption{Non-relativistic potentials energy curves for the $^2\Sigma^+_g$ and $^4\Sigma^+_g$ electronic states of the Sr$_2^+$ molecular ion.}
\label{fig:sg}
\begin{center}
\includegraphics[width=0.96\columnwidth]{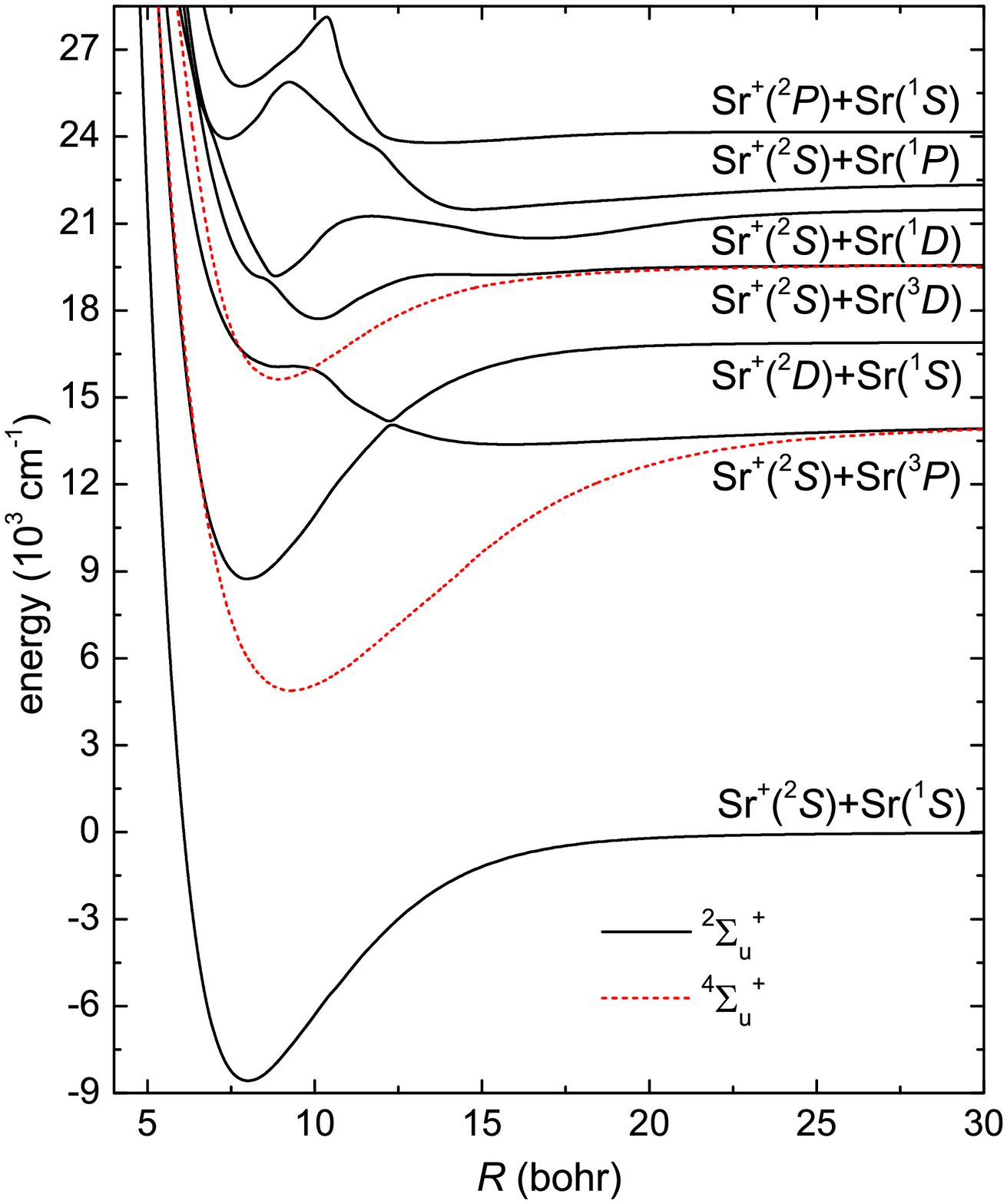}
\end{center}
\caption{Non-relativistic potentials energy curves for the $^2\Sigma^+_u$ and $^4\Sigma^+_u$ electronic states of the Sr$_2^+$ molecular ion.}
\label{fig:su}
\end{figure}
\begin{figure}[hptb!]
\begin{center}
\includegraphics[width=0.96\columnwidth]{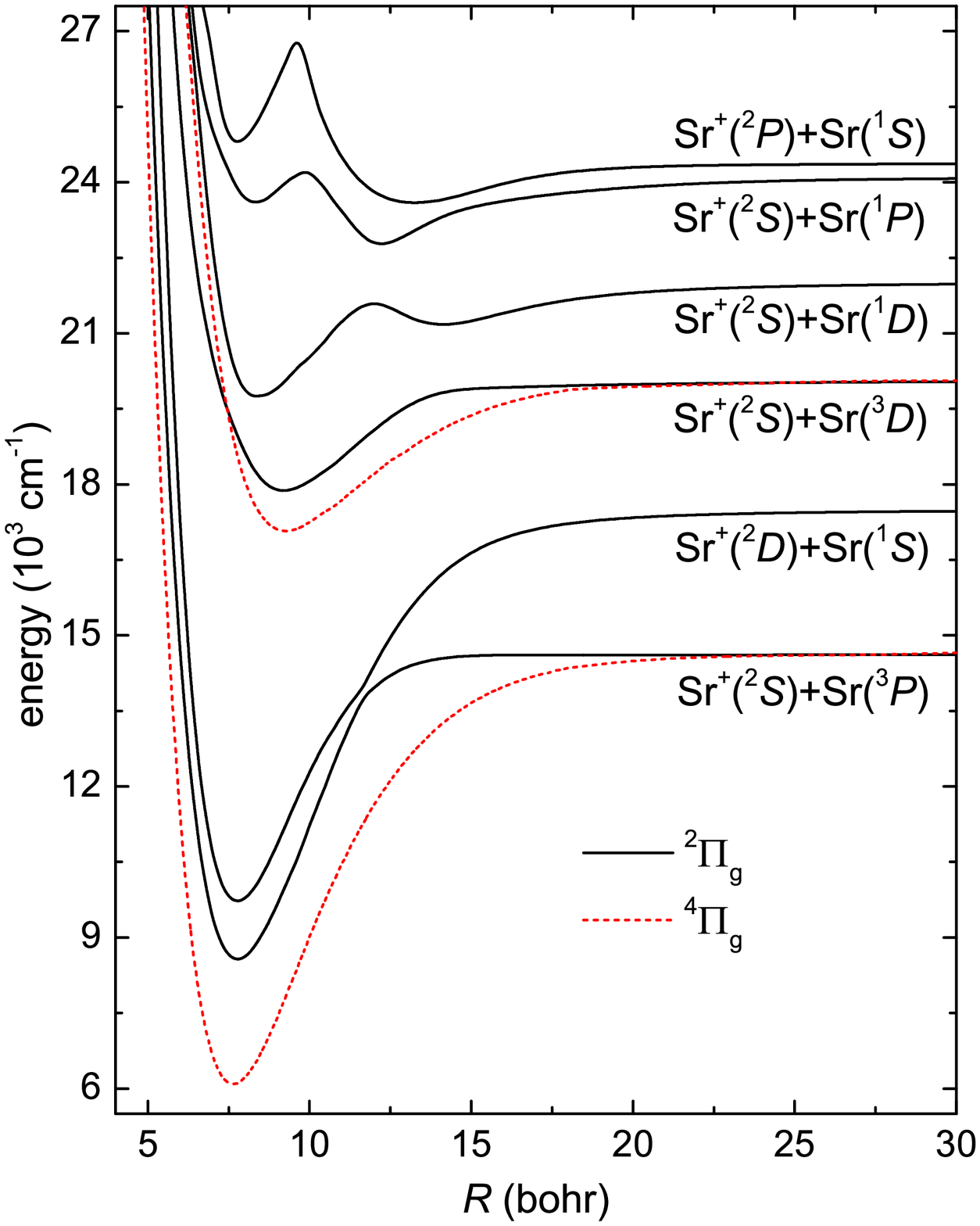}
\end{center}
\caption{Non-relativistic potentials energy curves for the $^2\Pi_g$ and $^4\Pi_g$ electronic states of the Sr$_2^+$ molecular ion.}
\label{fig:pg}
\begin{center}
\includegraphics[width=0.96\columnwidth]{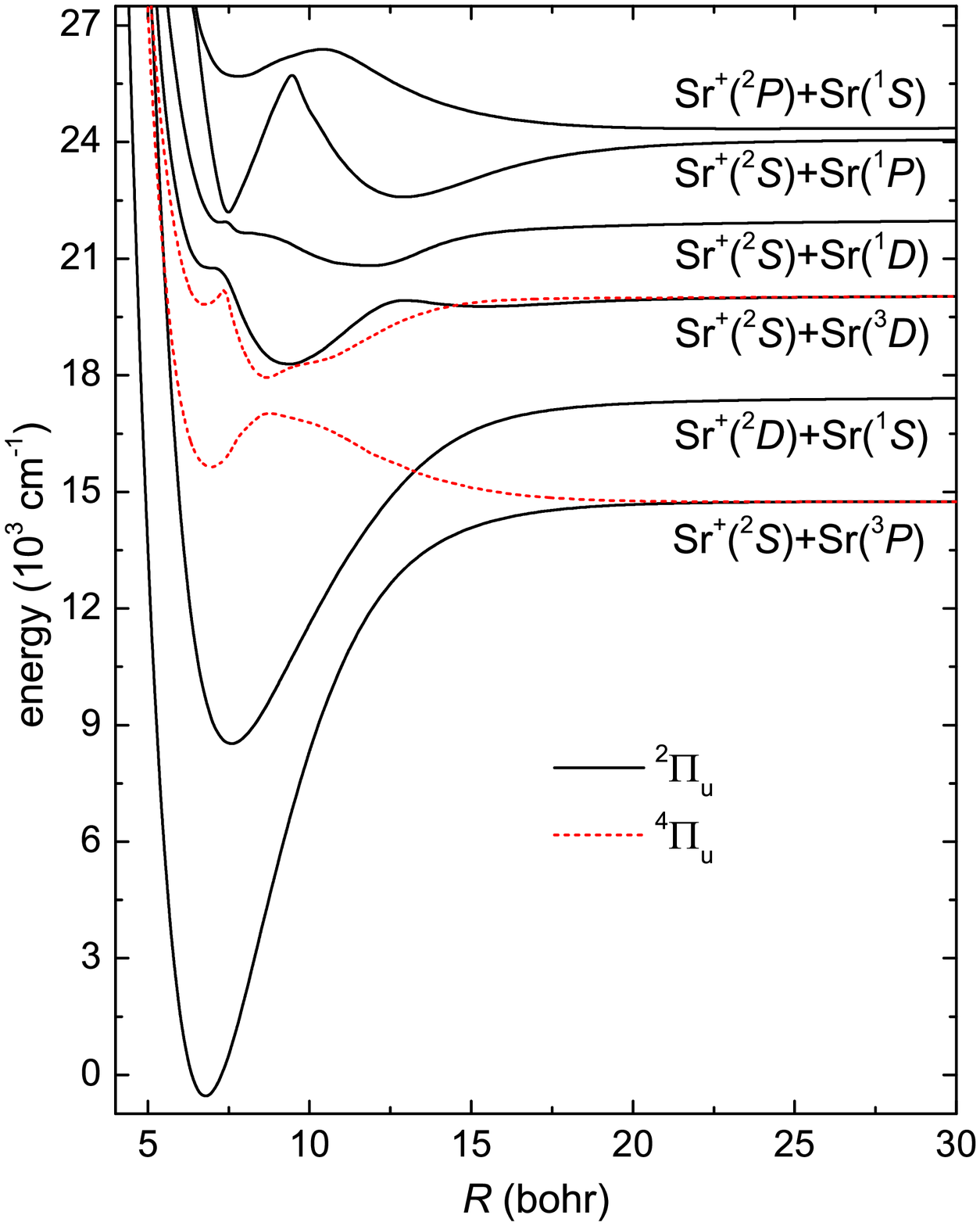}
\end{center}
\caption{Non-relativistic potentials energy curves for the $^2\Pi_u$ and $^4\Pi_u$ electronic states of the Sr$_2^+$ molecular ion.}
\label{fig:pu}
\end{figure}
\!\!of the short-range minimum, which can be reproduced by adding at least triples excitations.

\begin{table}[tb]
\caption{Spectroscopic characteristics of the Sr$_2^+$ molecular ion in $^2|\Lambda|_g$ electronic states: equilibrium bond length~$R_e$, well depth~$D_e$, harmonic constant~$\omega_e$, and rotational constant~$B_e$ obtained with the MRCISD and RCCSD(T) methods. \label{tab:dubletyg}} 
\begin{ruledtabular}
\begin{tabular}{lcrrrr}
State & Ref. & $R_e\,$(bohr) & $D_e\,$(cm$^{-1}$) &  $\omega_e\,$(cm$^{-1}$) & $B_e\,$(cm$^{-1}$) \\
\hline
\multicolumn{6}{c}{Sr$^+(^2S)$~+~Sr$(^1S)$} \\
$1^2\Sigma^+_g$  & MRCI &  9.23 & 678 & 37.5 & 0.0161 \\
$1^2\Sigma^+_g$  & CC & 10.03 & 78 & 23.7 & 0.0137 \\
\multicolumn{6}{c}{Sr$^+(^2S)$~+~Sr$(^3P)$} \\
$2^2\Sigma^+_g$  & MRCI  & 7.45 & 8027 & 104.4 & 0.0247  \\
$1^2\Pi_g$  & MRCI &  7.78 & 6044 & 105.2 & 0.0227 \\
$1^2\Pi_g$  & CC & 7.75 & 6041 & 81.5 & 0.0229  \\
\multicolumn{6}{c}{Sr$^+(^2D)$~+~Sr$(^1S)$} \\
$3^2\Sigma^+_g$  & MRCI & 8.46 & 5630 & 64.8 & 0.0192   \\
$2^2\Pi_g$  & MRCI & 7.77 & 7764 & 80.8 & 0.0228  \\
$1^2\Delta_g$  & MRCI & 7.20 & 10710 & 100.7 & 0.0265 \\
$1^2\Delta_g$  & CC & 7.23 & 9256 & 98.7 & 0.0241 \\
\multicolumn{6}{c}{Sr$^+(^2S)$~+~Sr$(^3D)$} \\
$4^2\Sigma^+_g$  & MRCI & 6.54 & 3013 & 422.7 & 0.0321  \\
$3^2\Pi_g$  & MRCI & 9.20 & 2182 & 48.0 & 0.0163 \\
$2^2\Delta_g$  & MRCI & 9.04 & 2425 & 48.7 & 0.0168 \\
\multicolumn{6}{c}{Sr$^+(^2S)$~+~Sr$(^1D)$} \\
$5^2\Sigma^+_g$  & MRCI &  10.60 & 1630 & 31.4 & 0.0122 \\
$4^2\Pi_g$  & MRCI & 8.36 & 2260 & 57.0 & 0.0197  \\
$3^2\Delta_g$  & MRCI & 13.51 & 702 & 33.6 & 0.0075 \\
\multicolumn{6}{c}{Sr$^+(^2S)$~+~Sr$(^1P)$} \\
$6^2\Sigma^+_g$  & MRCI  & 10.72 & 37 & 21.9 & 0.0120  \\
$5^2\Pi_g$  & MRCI &  12.21 & 1313 & 49.1 & 0.0092 \\
\multicolumn{6}{c}{Sr$^+(^2P)$~+~Sr$(^1S)$} \\
$7^2\Sigma^+_g$  & MRCI  & 14.70 & 891 & 15.6 & 0.0064  \\
$6^2\Pi_g$  & MRCI & 13.25 & 786 & 23.0 & 0.0078  \\
\end{tabular}
\end{ruledtabular}
\end{table}

\begin{table}[tb]
\caption{Spectroscopic characteristics of the Sr$_2^+$ molecular ion in $^2|\Lambda|_u$ electronic states: equilibrium bond length~$R_e$, well depth~$D_e$, harmonic constant~$\omega_e$, and rotational constant~$B_e$ obtained with the MRCISD and RCCSD(T) methods. \label{tab:dubletyu}} 
\begin{ruledtabular}
\begin{tabular}{lcrrrr}
State & Ref. & $R_e\,$(bohr) & $D_e\,$(cm$^{-1}$) &  $\omega_e\,$(cm$^{-1}$) & $B_e\,$(cm$^{-1}$) \\
\hline
\multicolumn{6}{c}{Sr$^+(^2S)$~+~Sr$(^1S)$} \\
$X^2\Sigma^+_u$  & MRCI & 7.99 & 8611 & 78.9 & 0.0215 \\
$X^2\Sigma^+_u$  & CC & 7.93 & 8576 &  79.9 & 0.0219 \\
\multicolumn{6}{c}{Sr$^+(^2S)$~+~Sr$(^3P)$} \\
$2^2\Sigma^+_u$  & MRCI  & 7.99 & 5345 & 79.3 & 0.0215  \\
$1^2\Pi_u$  & MRCI &  6.78 & 15290 & 93.2 & 0.0299 \\
$1^2\Pi_u$  & CC &  6.70 & 17030 & 125.9 & 0.0306 \\
\multicolumn{6}{c}{Sr$^+(^2D)$~+~Sr$(^1S)$} \\
$3^2\Sigma^+_u$  & MRCI & 12.22 & 2742 & 124.7 & 0.0092   \\
$2^2\Pi_u$  & MRCI & 7.59 & 8908 & 83.6 & 0.0239  \\
$1^2\Delta_u$  & MRCI & 8.10 & 7939 & 73.7 & 0.0210 \\
$1^2\Delta_u$  & CC &  7.89 & 6466 & 76.3 & 0.0215 \\
\multicolumn{6}{c}{Sr$^+(^2S)$~+~Sr$(^3D)$} \\
$4^2\Sigma^+_u$  & MRCI & 10.12 & 1876 & 62.3 & 0.0134  \\
$3^2\Pi_u$  & MRCI &  9.36 & 1761 & 44.4 & 0.0157 \\
$2^2\Delta_u$  & MRCI &  8.63 & 3575 & 52.2 & 0.0185\\
\multicolumn{6}{c}{Sr$^+(^2S)$~+~Sr$(^1D)$} \\
$5^2\Sigma^+_u$  & MRCI &  8.82 & 2373 & 129.1 & 0.0178 \\
$4^2\Pi_u$  & MRCI & 11.88 & 1179 & 23.2 & 0.0097  \\
$3^2\Delta_u$  & MRCI & 13.88 & 656 & 30.0 & 0.0071 \\
\multicolumn{6}{c}{Sr$^+(^2S)$~+~Sr$(^1P)$} \\
$6^2\Sigma^+_u$  & MRCI  & 14.71 & 945 & 26.6 & 0.0064  \\
$5^2\Pi_u$  & MRCI &  7.49 & 1900 & 225.0 & 0.0245 \\
\multicolumn{6}{c}{Sr$^+(^2P)$~+~Sr$(^1S)$} \\
$7^2\Sigma^+_u$  & MRCI  & 13.58 & 378 & 15.6 & 0.0075 \\
$6^2\Pi_u$  & MRCI & 23.50 & 28 & 2.2 & 0.0025  \\
\end{tabular}
\end{ruledtabular}
\end{table}

\begin{table}[tb]
\caption{Spectroscopic characteristics of the Sr$_2^+$ molecular ion in $^4|\Lambda|_g$ electronic states: equilibrium bond length~$R_e$, well depth~$D_e$, harmonic constant~$\omega_e$, and rotational constant~$B_e$ obtained with the MRCISD and RCCSD(T) methods. \label{tab:kwartetyg}} 
\begin{ruledtabular}
\begin{tabular}{lcrrrr}
State & Ref. & $R_e\,$(bohr) & $D_e\,$(cm$^{-1}$) &  $\omega_e\,$(cm$^{-1}$) & $B_e\,$(cm$^{-1}$) \\
\hline
\multicolumn{6}{c}{Sr$^+(^2S)$~+~Sr$(^3P)$} \\
$1^4\Sigma^+_g$  & MRCI & 22.59 & 315 & 6.52 & 0.0027\\
$1^4\Sigma^+_g$  & CC &  \multicolumn{4}{c}{repulsive}  \\
$1^4\Pi_g$  & MRCI & \multicolumn{4}{c}{repulsive}   \\
$1^4\Pi_g$  & CC & \multicolumn{4}{c}{repulsive}  \\
\multicolumn{6}{c}{Sr$^+(^2S)$~+~Sr$(^3D)$} \\
$2^4\Sigma^+_g$  & MRCI  & 13.80 & 116 & 20.6 & 0.0072  \\
$2^4\Pi_g$  & MRCI &  8.68 & 2105 & 81.4 & 0.0182 \\
$1^4\Delta_g$  & MRCI  & 6.28 & 10857 & 145.2 & 0.0348  \\
$1^4\Delta_g$  & CC & 6.26 & 10529 & 140.5 & 0.0327  \\
\end{tabular}
\end{ruledtabular}
\end{table}

\begin{table}[tb]
\caption{Spectroscopic characteristics of the Sr$_2^+$ molecular ion in $^4|\Lambda|_u$ electronic states: equilibrium bond length~$R_e$, well depth~$D_e$, harmonic constant~$\omega_e$, and rotational constant~$B_e$ obtained with the MRCISD and RCCSD(T) methods. \label{tab:kwartetyu}} 
\begin{ruledtabular}
\begin{tabular}{lcrrrr}
State & Ref. & $R_e\,$(bohr) & $D_e\,$(cm$^{-1}$) &  $\omega_e\,$(cm$^{-1}$) & $B_e\,$(cm$^{-1}$) \\
\hline
\multicolumn{6}{c}{Sr$^+(^2S)$~+~Sr$(^3P)$} \\
$1^4\Sigma^+_u$  & MRCI & 9.29 & 9194 & 54.9 & 0.0159\\
$1^4\Sigma^+_u$  & CC & 9.20 & 9490 & 49.9 & 0.0162  \\
$1^4\Pi_u$  & MRCI & 7.63 & 8524 & 77.0 & 0.0236  \\
$1^4\Pi_u$  & CC & 7.56 & 9004 & 79.7 & 0.0241 \\
\multicolumn{6}{c}{Sr$^+(^2S)$~+~Sr$(^3D)$} \\
$2^4\Sigma^+_u$  & MRCI  & 8.91 & 3962 & 53.4 & 0.0173  \\
$2^4\Pi_u$  & MRCI & 9.27 & 2985 & 51.2 & 0.0160 \\
$1^4\Delta_u$  & MRCI  & 8.67 & 4161 & 51.4 & 0.0183  \\
$1^4\Delta_u$  & CC & 8.64 & 3229 & 52.9 & 0.0184  \\
\end{tabular}
\end{ruledtabular}
\end{table}

\begin{figure}[tb]
\begin{center}
\includegraphics[width=0.96\columnwidth]{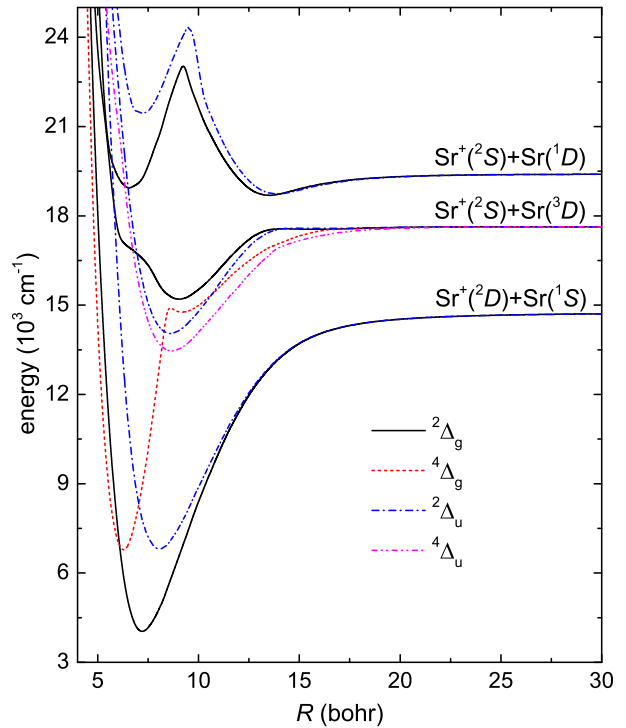}
\end{center}
\caption{Non-relativistic potentials energy curves for the $^2\Delta_g$, $^4\Delta_g$, $^2\Delta_u$, and $^4\Delta_u$ electronic states of the Sr$_2^+$ molecular ion.}
\label{fig:dgu}
\end{figure}

Potential energy curves of the ground and excited doublet and quartet electronic states of $\Sigma_g^+$, $\Sigma_u^+$, $\Pi_g$, $\Pi_u$, $\Delta_g$, and $\Delta_u$ symmetries are presented in Figs.~\ref{fig:sg}-\ref{fig:dgu}. Their spectroscopic characteristics including equilibrium bond lengths~$R_e$, well depths~$D_e$, harmonic constants~$\omega_e$, and rotational constants~$B_e$ are gathered in Tables~\ref{tab:dubletyg}-\ref{tab:kwartetyu}. Based on the MRCISD results, we can identify that the $1^2\Pi_u$ state is the most strongly bound with $D_e=15290\,$cm$^{-1}$, followed by $1^4\Delta_g$ and $1^2\Delta_g$ with dissociation energies of 10857$\,$cm$^{-1}$ and 10710$\,$cm$^{-1}$, respectively. These three electronic states also exhibit the shortest equilibrium bond lengths of 6.78$\,$bohr, 6.28$\,$bohr, and 7.20$\,$bohr, respectively. On the other hand, the most weakly bound states are $6^2\Pi_u$, $6^2\Sigma_g^+$, and $2^4\Sigma_g^+$, for which the calculated dissociation energies are equal to 28$\,$cm$^{-1}$, 37$\,$cm$^{-1}$, and 116$\,$cm$^{-1}$, respectively. The largest equilibrium bond length is predicted for the $6^2\Pi_u$ state at 23.5$\,$bohr, followed by 22.6~bohr for the $2^4\Sigma_g^+$ state and 14.7$\,$bohr for the $6^2\Sigma_u^+$ state. We expect relative errors for deeply-bound electronic states to be significantly smaller than for weakly-bound ones.

By analyzing the pattern of the potential energy curves, we find that many of them display avoided crossings, suggesting strong radial non-adiabatic couplings between involved electronic states. We further observe a general tendency that lower-lying potential energy curves show a smooth behavior with well defined minima, while higher-lying states display perturbations, mostly in the form of avoided crossings due to the interaction with other electronic states of the same symmetry that result from closely-lying atomic thresholds. At high energies the density of electronic states becomes prohibitively large for MRCISD calculations, and for this reason we restricted  our analysis to the seven lowest-lying dissociation limits, although the highest-lying calculated states show avoided crossing with yet higher asymptotes.

\begin{figure*}[tb]
\begin{center}
\includegraphics[width=0.95\textwidth]{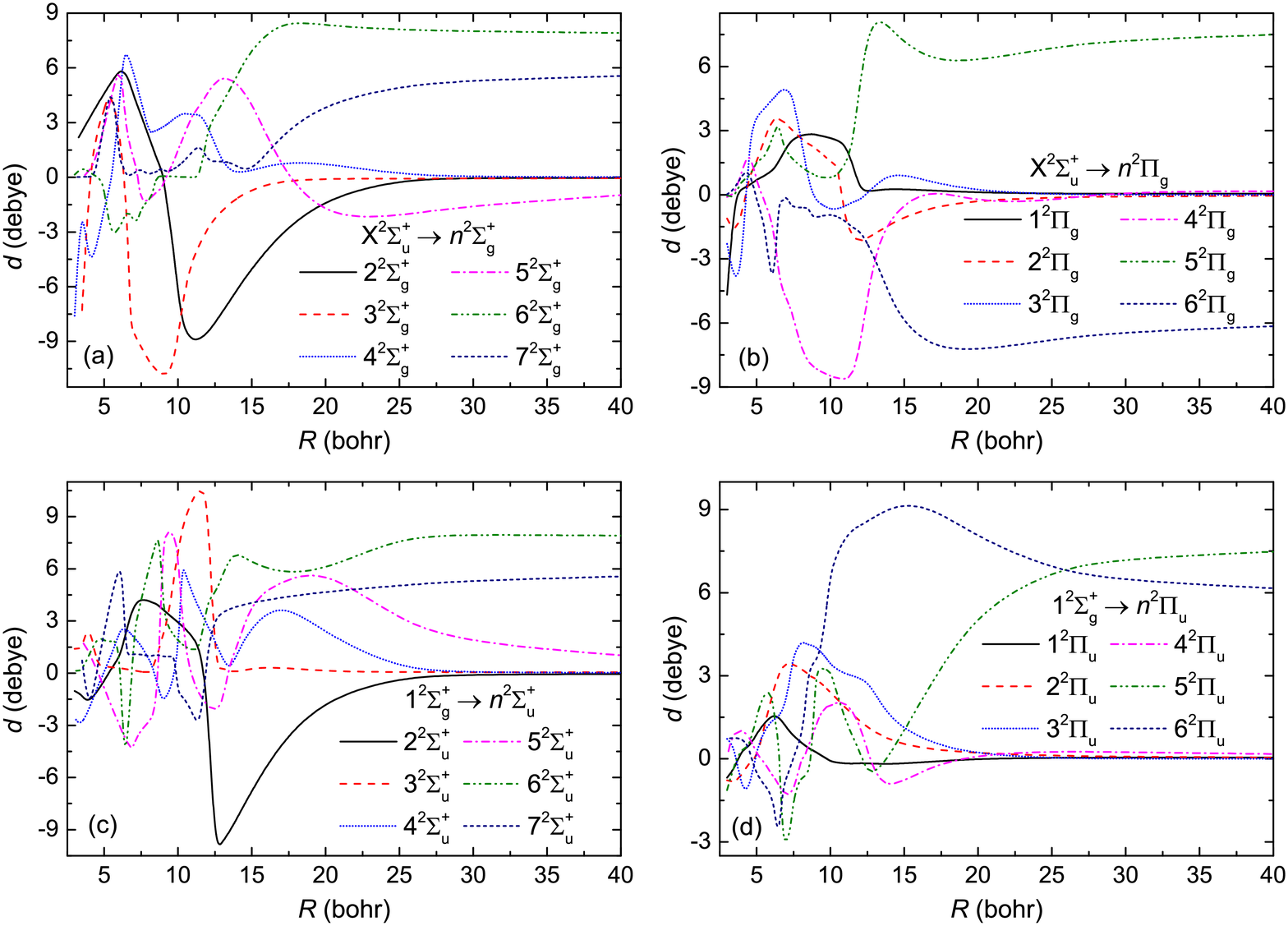}
\end{center}
\caption{Electric dipole transition moments between (a) the $X^2\Sigma^+_u$ ground and $^2\Sigma^+_g$ exited states, (b) the $X^2\Sigma^+_u$ ground and $^2\Pi_g$ exited states, (c) the $1^2\Sigma^+_g$ lowest excited and $^2\Sigma^+_u$ exited states, and (d) the $1^2\Sigma^+_g$ lowest excited and $^2\Sigma^+_u$ exited states of the Sr$_2^+$ molecular ion.}
\label{fig:dip}
\end{figure*}

The accuracy of our molecular calculations could be estimated by a comparison with the available experimental results. In the present case, we compare the obtained dissociation energy for the ground state of the Sr$_2^+$ ion with the experimental value of $8800(130)\,$cm$^{-1}$~\cite{dugourdCPL92}. The MRCISD value of 8611$\,$cm$^{−1}$ and the RCCSD(T) one of 8576$\,$cm$^{−1}$ correspond to the errors of 2.2\% and 2.6\%, respectively. The MRCISD+Q result agrees perfectly with the experimental value but the Davidson correction may give unreasonable results for excited states in the vicinity of avoided crossings. Unfortunately, no experimental excitation energies or dissociation energies for exited states of the Sr$_2^+$ molecular ion are available. Therefore, our estimation of possible error margins for excited electronic states can only be based on the ability of our approach to reproduce atomic and ionic properties as described in Sec.~\ref{sec:theory}, the convergence of results for the two lowest states presented above, and  the reproduction of exited atomic limits in molecular calculations.  

The obtained excitation energies at atomic dissociation limits with the MRCISD and RCCSD(T) methods together with their experimental counterparts are presented in Table~\ref{tab:thresholds}. An analysis of these data shows that the excitation energies from the ground state to all but one excited states are reproduced by the MRCISD and RCCSD(T) methods within 5\%. The notable exception is the excitation of the Sr$_2^+$ ion to the excited states corresponding to the Sr$^+(^2D)$+Sr$(^1S)$ dissociation limit. A possible explanation of this behavior of both theoretical methods is the unsaturation of the employed basis set with $d$-type orbitals, combined with an extreme proximity of the lower-lying asymptote, which should be separated from the considered limit by 40$\,$cm$^{-1}$ only. As a result, the absolute error of this excitation energy is equal to around 2000$\,$cm$^{-1}$ for MRCISD and 1100$\,$cm$^{-1}$ for RCCSD(T), that corresponds to 14\% and 7.5\% relative errors, respectively.

Based on the above analysis and our previous experience with \textit{ab initio} calculations for similar systems, we estimate the uncertainty of the calculated interactions energies to be around 5\% for the low-lying deeply bound electronic states unaffected by avoided-crossings, whereas for high-lying states disturbed by avoided-crossings and high density of states  to be in the range of 10\% to 20\%, or even larger for weakly bound highly exited states.

It is also worth to mention, that to obtain presented converged results and smooth potential energy curves a special numerical treatment, optimized separately for all electronic symmetries has been needed including restarting calculations from different geometries and configurations, and properly adjusting  numerical thresholds.

Summarizing this part, we have shown that the alkaline-earth-metal dimers, including Sr$_2^+$ may pose a considerable challenge even for the state-of-the-art \textit{ab initio} theoretical methods. Among the possible reasons, one can mention a complicated multireference electronic structure of these species, which exhibit closely lying excited electronic states with strong radial non-adiabatic couplings. Therefore, the theoretical description of some states of the Sr$_2^+$ molecular ion may suffer from a decreased accuracy. Nonetheless, the presented results contribute significantly to the yet unknown energetic structure of this interesting molecular ion. Additionally, a demanding character of the Sr$_2^+$ ion can be treated as an opportunity to benchmark a range of theoretical methods, which are often considered as methods of choice for similar systems. A comparison of their performance in the description of the ground and some excited states has been an important aspect of our study.

\subsection{Electric dipole transition moments}

\begin{figure*}[tb]
\begin{center}
\includegraphics[width=\columnwidth]{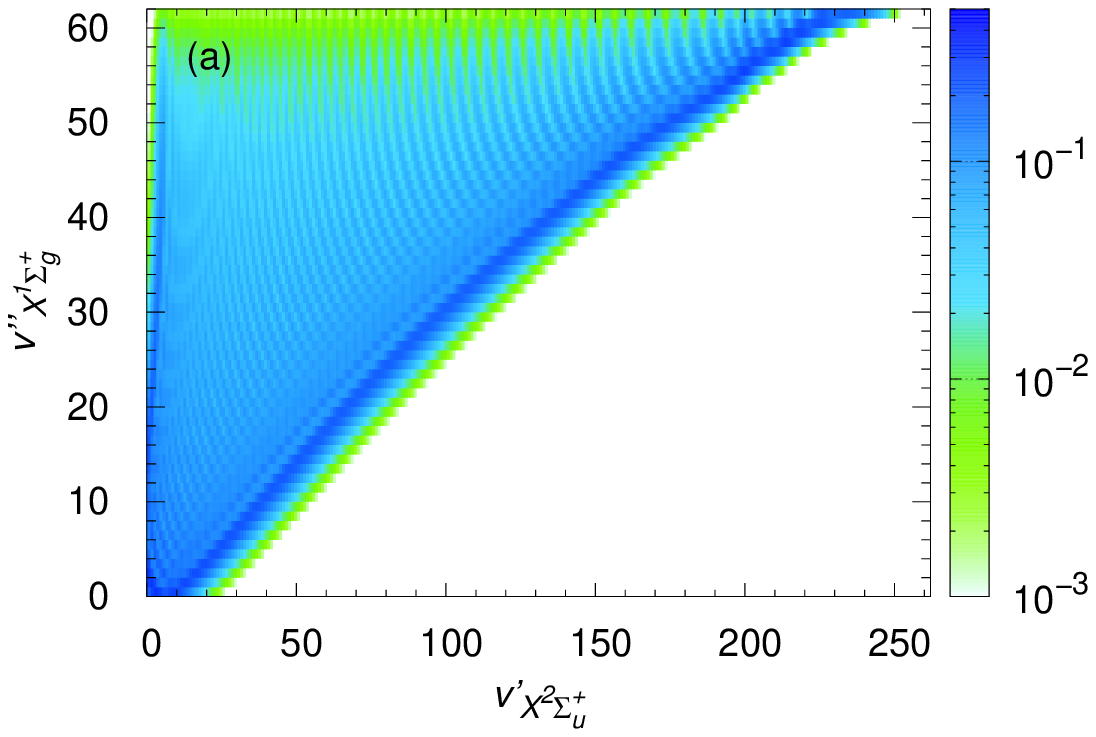}
\includegraphics[width=\columnwidth]{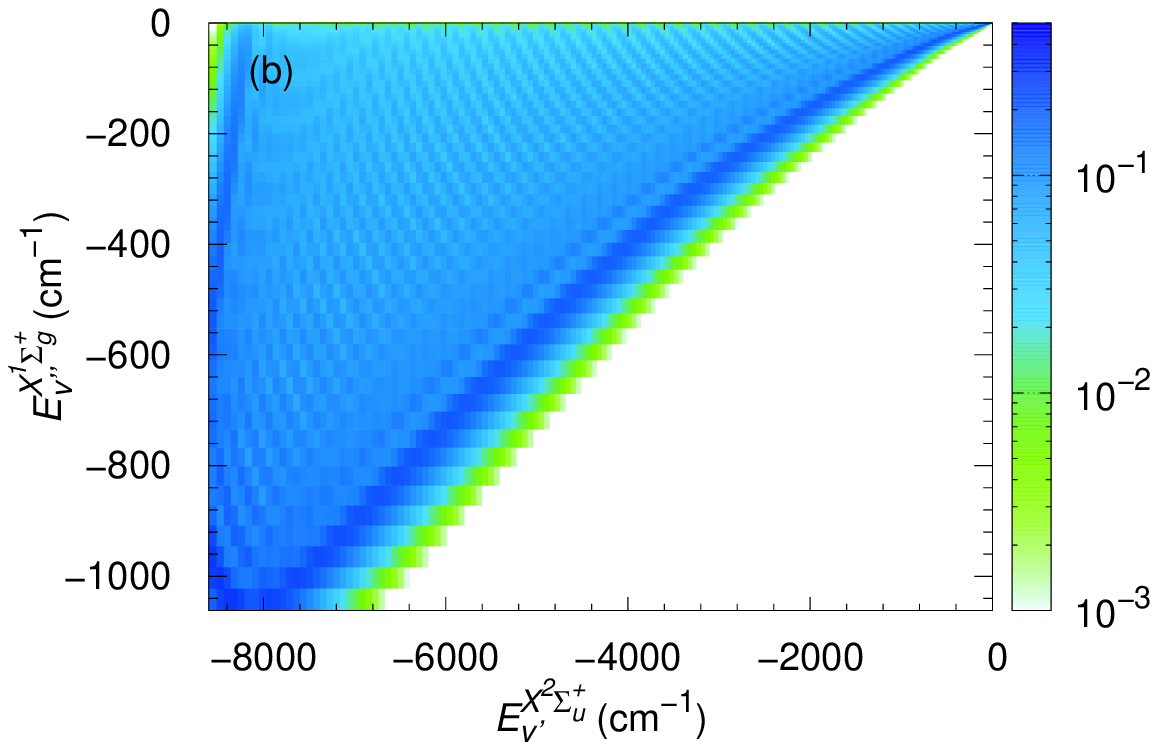}
\includegraphics[width=\columnwidth]{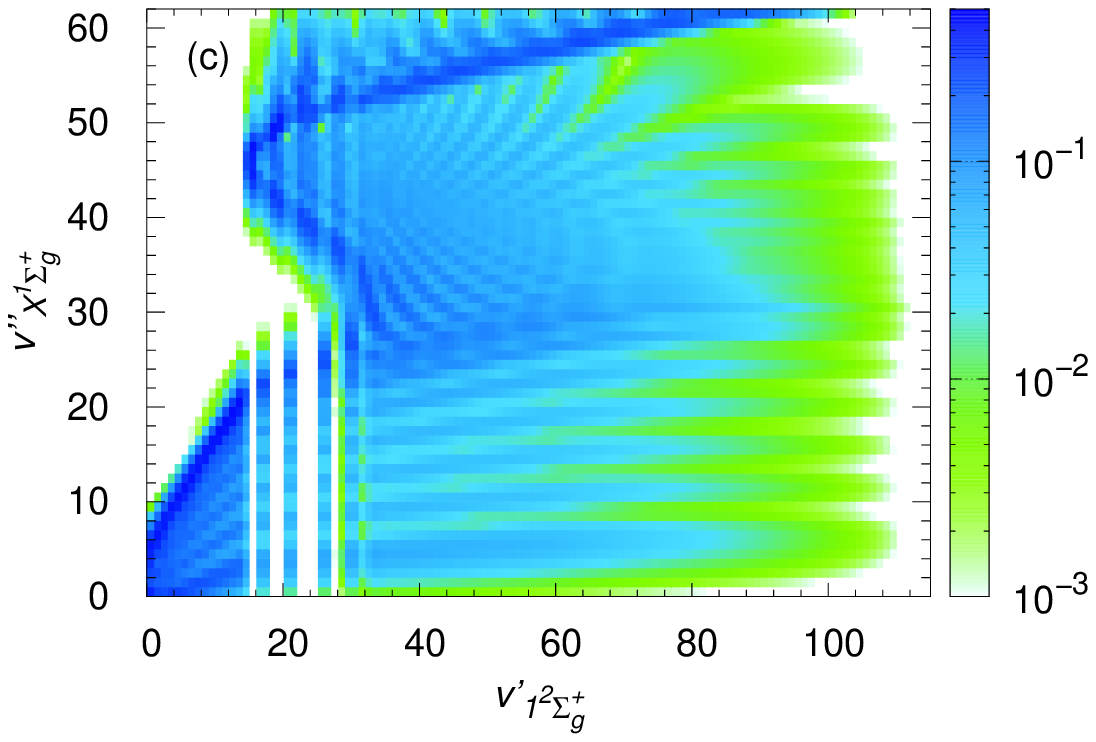}
\includegraphics[width=\columnwidth]{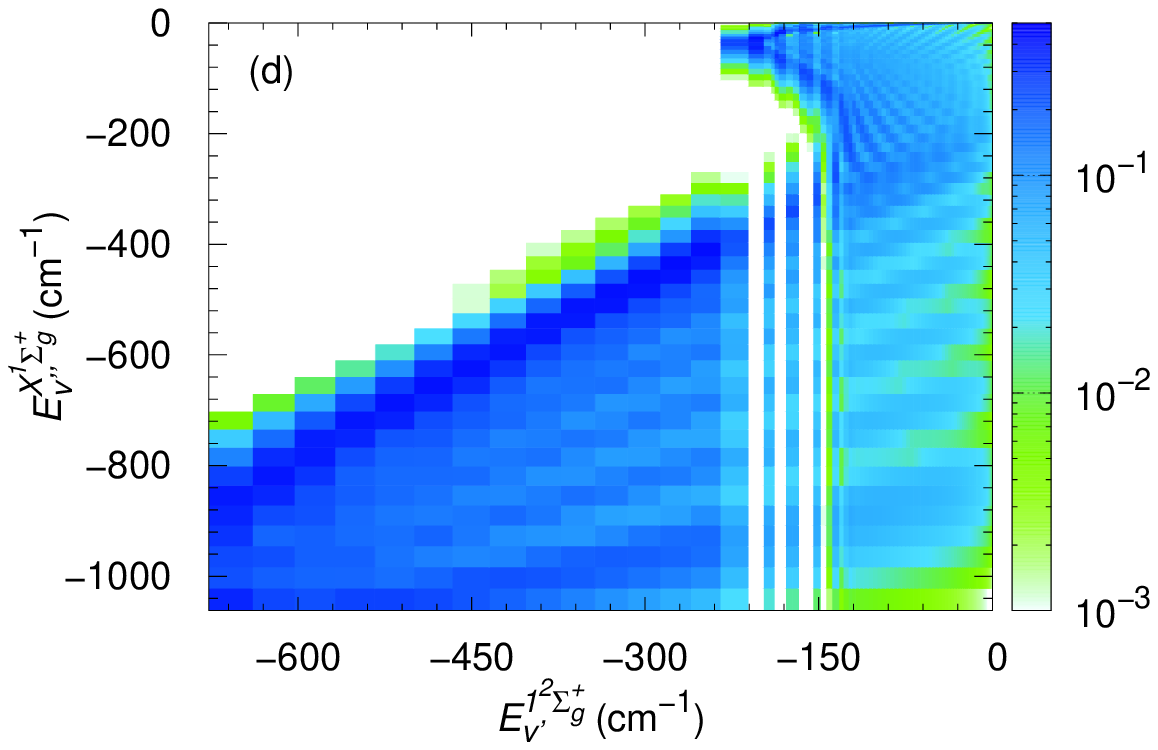}
\end{center}
\caption{Franck-Condon factors between vibrational levels of the $X^1\Sigma^+_g$ ground electronic states of the $^{88}$Sr$_2$ molecule and vibrational levels of (a),(b) the $X^2\Sigma^+_u$ and (c),(d) the $1^2\Sigma^+_g$ electronic states of the $^{88}$Sr$_2^+$ molecular ion  as a function of the vibrational quantum numbers (a),(c) and vibrational energies (b),(d).}
\label{fig:FCF}
\end{figure*}

The electric dipole transition moments, necessary for a full characterization of the molecular spectra, are calculated within the MRCISD method and are presented in Fig.~\ref{fig:dip} for transitions from the $X^2\Sigma_u^+$ and $1^2\Sigma_g^+$ electronic states to all possible excited states. 
The strongest transition moments reach around 9$\,$debye, e.g.~from the equilibrium interatomic distance in the $X^2\Sigma^+_u$ ground state to the $3^2\Sigma^+_g$ state as well as to the $4^2\Pi_g$ state. Some transition moments are clearly smaller for the equilibrium interatomic distance of the ground state Sr$_2^+$, suggesting that the corresponding line intensities in the spectra will be weaker. Depending on the electronic configuration of atomic limits, transitions may be dipole-allowed or dipole-forbidden for large interatomic distances. The electric dipole transition moment curves exhibit high variability and strong dependence on the interatomic distance at shorter ranges due interaction-induced modifications of the underlying electronic structure. Additionally, the complexity of the electric dipole transition moment curves may be attributed to the distortion of the excited states by many avoided crossings.

\subsection{Franck-Condon factors}

Cold molecular ions can be formed by photoassociation of laser-cooled Sr$^+$ ions immersed into an ultracold gas of Sr atoms or by ionization of ultracold Sr$_2$ molecules. Following recent successes in the production of ultracold gases of Sr$_2$ molecules, we calculate Franck-Condon factors governing the photoionization of ground-state $^{88}$Sr$_2$ molecules into the two lowest electronic states of $^{88}$Sr$_2^+$ molecular ions, while the calculation of the photoionization cross sections is out of the scope of this paper. The Schr\"odinger equation for nuclear motion is solved as implemented and described in Ref.~\cite{TomzaPRA15a}.

Fig.~\ref{fig:FCF} presents calculated Franck-Condon factors (FCFs) between vibrational levels of the $X^1\Sigma^+_g$ ground electronic state of the $^{88}$Sr$_2$ molecule and vibrational levels of the $X^2\Sigma^+_u$ and $1^2\Sigma^+_g$ electronic states of the $^{88}$Sr$_2^+$ molecular ion as a function of the vibrational quantum numbers and vibrational energies. The $X^1\Sigma^+_g$ ground state of the $^{88}$Sr$_2$ molecule supports 63 rotationless vibrational levels~\cite{SteinPRA08}, while the $X^2\Sigma^+_u$ and $1^2\Sigma^+_g$ electronic states of the $^{88}$Sr$_2^+$ molecular ion supports 263 and 116 vibrational levels, respectively. Because the ground states of Sr$_2$ and Sr$_2^+$ have similar equilibrium bond lengths, FCFs for the $X^1\Sigma^+_g$-$X^2\Sigma^+_u$ transition are regular and have a pronounced diagonal band, with the largest value of 0.38 for $v''_{X^1\Sigma^+_g}=0\to v'_{X^2\Sigma^+_u}=5$ overlap. The asymmetry visible in Fig.~\ref{fig:FCF}(a),(b) results from a significant difference in well depths of the neutral and ionic grounds states. Deeply bound Sr$_2^+$ molecular ions in the $X^2\Sigma^+_u$ state can be obtained by the ionization of deeply bound Sr$_2$ molecules, however noticeable probability can be expected for the ionization of weakly bound molecules, too. The ionization of weakly bound Sr$_2$ molecules into weakly bound molecular ions can also be expected. FCFs for the $X^1\Sigma^+_g$-$1^2\Sigma^+_g$ transition in Fig.~\ref{fig:FCF}(c),(d) are less regular due to the double well structure of the $1^2\Sigma^+_g$ state. Levels of the short-range well of the $1^2\Sigma^+_g$ state have a relatively good overlap with levels of the $X^1\Sigma^+_g$ state with the largest value of 0.62 for $v''_{X^1\Sigma^+_g}=21\to v'_{1^2\Sigma^+_g}=14$ transition. The vertical lines visible in Fig.~\ref{fig:FCF}(c),(d) are related with the FCFs to levels localized in the long-range well of the $1^2\Sigma^+_g$ state, which do not have noticeable overlap with deeply bound levels of the $X^1\Sigma^+_g$ state. Deeply bound Sr$_2^+$ molecular ions in the $1^2\Sigma^+_g$ state can be obtained by the ionization of deeply bound Sr$_2$ molecules, only. Interestingly, Sr$_2^+$ molecular ions localized in the long-range well of the $1^2\Sigma^+_g$ state can be produced by the ionization of weakly bound Sr$_2$ molecules.

\section{Summary and conclusions}
\label{sec:summary}

Cold molecular ions consisting of alkali and alkaline-earth atoms are amongst the most promising systems for interesting applications ranging from quantum simulations and precision measurements to controlled chemical reactions. In comparison to neutral molecules, molecular ions offer better preparation, trapping and detection possibilities. At the same time, while neutral alkaline-earth dimers, including several recent works on Sr$_2$, have already been intensively studied, the lack of spectroscopic data for ionic alkaline-earth dimers may limit possible experimental efforts.

In the present work, therefore, we have calculated and characterized the ground and 41~excited electronic states of the Sr$_2^+$ molecular ion and electric dipole transition moments between them. We have employed a range of \textit{ab initio} quantum-chemical methods including the coupled cluster and configuration interaction ones with small-core relativistic energy-consistent pseudopotentials and large Gaussian basis set. While low-lying potential energy curves show a smooth behavior with well defined minima, the higher-lying states display perturbations, mostly in the form of avoided crossings. We have characterized and benchmarked two lowest electronic states of the Sr$_2^+$ molecular ion with a range of computational techniques and we have observed that the Sr$_2^+$ molecular ion despite of its apparently simple structure with three valence electrons only is a challenging system even for state-of-the-art \textit{ab initio} theoretical methods due to its multireference nature and high density of excited electronic states.  Finally, we have calculated and analyzed Franck-Condon factors governing the photoionization of ground-state Sr$_2$ molecules into the two lowest electronic states of Sr$_2^+$ molecular ions. We have shown that the formation of both weakly and deeply bound Sr$_2^+$ molecular ions using photoionization of weakly and deeply bound Sr$_2$ molecules should be feasible. 

As no experimental data is available for the Sr$_2^+$ molecular ion excitation energies, our results can be useful for guiding future spectroscopic measurements as well as for the formation of cold Sr$_2^+$ dimers by means of photoionization of neutral Sr$_2$ molecules from molecular beam or ultracold gas. In the future, we plan to use the calculated spectrum of the Sr$_2^+$ molecular ion in theoretical studies of an atomic ion immersed in ultracold gas of Sr$_2$ molecules where the charge transfer process is possible.
\

\section*{Supplementary Material}
Supplementary materials contain all potential energy curves and Franck-Condon factors in numerical form as plotted in all figures presented in the manuscript.

\begin{acknowledgments}
Financial support from the National Science Centre Poland (2015/19/D/ST4/02173 and 2016/23/B/ST4/03231) is gratefully acknowledged.
The computational part of this research has been partially supported by the PL-Grid Infrastructure.
\end{acknowledgments}

\bibliography{Sr2+}

\end{document}